\documentstyle[12pt]{article}
\begin{document}

\begin{center} {\huge {\bf Modelling and Computer Simulation of an Insurance 
Policy: A Search for Maximum Profit}} \end{center}

\vskip 1.5 cm

\begin{center} {\it  Muktish Acharyya$^1$ and Ajanta Bhowal Acharyya$^2$}\end{center}

\vskip 0.5cm

\begin{center} {\it $^1$Department of Physics, Krishnanagar Government College,}\\
 {\it PO-Krishnanagar, Dist-Nadia, PIN-741101, WB, India}\\
 {\it E-mail: muktish@vsnl.net}\end{center}

\vskip 0.5cm

\begin{center} {\it $^2$Department of Physics, Hooghly Mohsin College,}\\
{\it PO-Chinsurah, Dist-Hooghly, PIN-712101, WB, India}\end{center}

\vskip 2cm

\noindent {Abstract:}{\small We have developed a model for a life-insurance 
policy.
In this model, the net gain is calculated by computer simulation
for a particular type of lifetime distribution function. We observed that the 
net gain becomes maximum for a particular value of
upper age for last premium. This paper is dedicated to 
Professor Dietrich Stauffer on the occassion of his 60th birthday.}

\vskip 0.5cm


\vskip 1cm

\noindent {\bf 1. Introduction}

The subject {\it econophysics}\cite{ds} has attracted much
attention of statistical physicists due to its
rich variety, applicability and utility in our socio-economic life. 
A great effort has already been made by the 
statistical physicists to develop some 
mathematical model for economic growth, to predict the high time for share and
stock market, to study the fluctuations in financial index etc.\cite{rev} 
The active research is still going on in this field.  
The dependence of profit and loss of a marketing company
on various parameters is also an interesting and important 
area of research in the field of {\it econophysics}.
The insurance company (IC) is also a marketing company which always seeks for 
profit by providing various types of insurances for the common
people. 
However, we did not find any work (in {\it econophysics}) to develop a model for 
any kind of insurance policy and to calculate the profit as a function of various
parameters. This type of study does not claim to be new to the insurance science but intends to
introduce the {\it econophysicists} to this field by a simple example.
Let us confine our attention to the life insurance only. The
terms and conditions for a life insurance policy can be described simply as 
follows: one has to pay some amounts of money (called premium) yearly upto a certain
age. If the person dies before the end of the policy his/her
nominee will get some
amounts of money (promised by IC) from the insurance company 
irrespective of the total amount of
premium paid to IC by that person. 
Sometimes this is called sum assured value (SAV).
Thus IC has some apparent loss since the person
dies much before the age of last premium. If the person survives upto the end
of this policy (s)he will also get some amount of money. 
This amount may be called the payment on maturity (POM).
In this way IC provides insurance for life as well 
as acts as an investment sector
for the common people. 
But naturally everybody will not die before the
end of this policy. The deaths of people also obey some natural rule
which is reflected in the statistical distribution of lifetimes
which is called the rate of survival by demographers. The 
% sociologist ==> demographer
probability of survival upto an age is a nonmonotonic function having
a maximum. That means, most of the persons in the society will die at that particular
age due to various reasons. 
IC makes profit from the persons who survive at least upto the
age of last premium. 
In this way, the IC
exploits the nature of lifetime distribution and tries to get maximum profit.

In this paper we propose a model of a particular (life) insurance 
policy and study it by computer simulation.
We calculated the net profit and its variation 
with the upper age (age for last premium of a person who is alive)
of the insurance policy. Interestingly, we found, the existence of an upper
age of last premium for which the profit will be maximum. Accordingly IC has to
set the upper age of last premium to maximise the profit. 

We have orgainsed the paper as follows:
In the next section we describe the model and its computer 
simulation method, in the third section we have shown the
numerical results and the paper ends with a concluding remarks
given in the last section. 

\vskip 1cm

\noindent {\large {\bf 2. The description of the model and simulation method}}

Before coming to the detailed description of our proposed model for 
life insurance, let us first explain in a bit detail
what a life insurance is. A person having life insurance policy, has to 
pay some amount of money (fixed by the
insurance company) every year starting from age $T_i$ (fixed by the 
insurance company) until the age becomes  
$T_f$ (also fixed by the insurance company). 
Here, it may be noted that this $T_i$ is the minimum age to have insurance. 
One may take insurance after this age but
we are considering here only those persons who took insurance at the age $T_i$.
After
paying the last premium he/she (if still alive) will get back some amount of 
money (with some interest), which is called payment on maturity (POM). 
But, if any  
person dies before the age ($T_f$) of paying the 
last premium, his/her nominee will get the sum assured value (SAV)
% also omitted
(amount promised by IC). 
Now if we think from the IC's point of view, every year
IC is getting premiums from different persons and paying 
the SAV to those who are dying in that particular year.
So, there is a competition between the total 
premiums collection and the payment of sum 
assured value (SAV) due to death. The difference of 
these two amounts is the yearly profit (YP) of IC at 
that particular year. 

Now, firstly, IC will invest 
these amounts of money (YP) every year
and at the end of the policy it will
get some amount with interest. 
Secondly, at the end of the policy, IC has to pay the payment on maturity
(POM) to the living persons. 
The net gain of IC, after the end of this policy, is simply the difference of these two amounts. 

We design the model in the following way: 
suppose there are $N$ number of persons of 
same age having a particular policy for the
insurance of their life. The distribution
of their lifetimes $T_L$ is given by
\begin{equation}
P(T_L) = A T_L^a {\rm exp} (-{T_L^b}/c)
\end{equation}
%The distribution of lifetime is taken as realistic as possible. 
% Above line replaced by following sentence
It should be mentioned here that the above distribution of lifetimes
differs from that obtainable from the famous
Gompertz (1825) law which states that the probability
that an individual, now aged $x$, would survive to age $x+1$ is ${\rm exp}[-\int_x^{x+1} BC^x dx] = e^{-k^x}$.
We compare the chosen distribution of lifetimes (eqn-1) with that obtained from
available data \cite{mort} given in table-I. 
For a first estimate this assumption (eqn-1) may be good though precise estimates
require an extrapolation of actualy observed mortalities into the future.
Here, the most probable lifetime is set at 75 years. To make these
most probable lifetime at $T_m = ({ac \over b})^{1 \over b} =75$ 
years we have taken $a=4.0$, $b=7.5$ and $c=2.16\times10^{14}$. 
Suppose, an insurance company
came to the market to insure the life of each of these $N$ persons. 
We are considering a particular policy, for which,
the minimum age, to have the policy, is taken as $T_i$=25 years.
The conditions of the policy are as follows: 
(i) every person having policy, will have to pay a yearly premium of
amount $P_r$ (in some arbitrary unit of currency). 
The duration of such an insurance policy 
is $n_t=T_f-T_i$ years. That means the last premium
has to be paid by the person (if alive) in his/her age of $T_f$ years. 
So, the person (if alive) is giving $P_r(T_f-T_i)$
amount of money to the insurance company (IC). This total 
amount obtained from personal premium is $R$. We fixed the value
$R = 1$ (in arbitrary currency unit). 
Here, throughout this paper, all the amounts of money are calculated in the unit
of $R$.
The amount of 
yearly premium ($P_r$) is calculated as $P_r = R/(T_f-T_i)$.
It is expected that the POM will be higher than $R$.
But, how the value of POM ($P_m$) will be decided ? In this model, we have
considered this as follows: 
Suppose, the rate of simple interest
given by IC is $r_1$. That means if 
someone invests the amount $P_r$ yearly (as premium)
for $n_t$ years to IC, he/she will receive finally
the amount $P_m=P_rn_t(1+{{r_1 \times (n_t+1)} \over 200})$.  
(ii) whereas if any policy holder dies at the age $T_d (<T_f$), the IC will return the person the amount SAV ($S_a$). 
It may be noted that $S_a$ is set slightly higher than $P_m$ which is generally done by IC.
In this model, we have taken $S_a=1.1\times P_m$.

We have studied the model taking $N$ different persons of same age and having 
the same type of life insurance policy. The lifetimes of $N$ persons
are distributed according to the relation (1). We set the initial and the final 
ages of this policy at $T_i$ and $T_f$  years
respectively. It (same $T_i$) means that all the persons in this set of peoples 
started their policy at their age of 25 years. Now the time evolution starts. The
unit of time is in year. We first counted the number of persons $N_L$ living in 
time $T$. Calculated the total premium collected, i.e.,
$N_L P_r$. At every time step (year), we also checked the number of persons who
died $N_d$, (from the distribution of lifetime)
in that particular year, to calculate the amount $N_d S_a$ which has to
be returned by the IC to the nominee of the persons died. Then calculated the gain 
in that year which is $G^{\prime} = N_L P_r - N_d S_a$. 
IC will invest this amount of money every year (with simple rate of $r_2$ percent
interest) to make the profit. 
For that particular year, the profit is calculated as $G(T) = G^{\prime}(1+r_2 (T_f-T) 0.01)$.
At the end of the year $T_f$ the total
profit of IC is calculated as $\int G(T) dT$. At this stage we calculated the 
number of persons still alive ($N_a$) and subtracted the amount $N_a P_m$ to
calculate the net gain $G_n = \int G(T) dT - N_a P_m$ of IC for this particular type of policy.
We finally calculated, the net gain per head, i.e., 
$G_n/N$ and studied it as a function of $T_f$.  

\vskip 1 cm

\noindent {\large {\bf 3. Numerical results}}

We studied the model described above for $N = 100000$ taking $r_1 = 2.0$ and $r_2 = 8.0$ 
The amount is taken in any arbitrary
currency units. $R = 1.0$, $T_i = 25$ years. The results of numerical simulation are 
shown in fig-1 and fig-2. Figure-1 shows the distribution
of lifetimes. Here it may be noted that the most probable lifetime was taken 75 years. 

%%%%%%%%%%%%%%%%%%%%%%%%%%%%%%%%
\vskip 1 cm

\begin{center}{\bf Table-I}\end{center}

\vskip 0.5 cm

\begin{center}
%%%%  TABLE %%%%%%%%%%%%%%%%%%%%%%%%%%%%%%%%
\begin{tabular}{|c|c|c|c|c|}
\hline
Age group & Number of deaths \\
\hline
1-4 years & 5287  \\
\hline
5-14 years & 7647  \\
\hline
15-24 years & 30,909  \\
\hline
25-44 years & 1,30,886   \\
\hline
45-64 years & 3,92,916  \\
\hline
65-74 years & 4,53,411 \\
\hline
75-84 years & 6,99,143 \\
\hline
84 above   & 6,46,333 \\
\hline

\end{tabular}
%%%%  END TABLE %%%%%%%%%%%%%%%%%%%%%%%%%%%%

\vskip 0.5 cm

\noindent {Table-I. The available data for mortality rates collected from ref. \cite{mort}.}

\end{center}
%%%%%%%%%%%%%%%%%%%%%%%%%%%%%%%%

%%% FIGURE-1
% GNUPLOT: LaTeX picture
\setlength{\unitlength}{0.240900pt}
\ifx\plotpoint\undefined\newsavebox{\plotpoint}\fi
\sbox{\plotpoint}{\rule[-0.200pt]{0.400pt}{0.400pt}}%
\begin{picture}(1500,900)(0,0)
\font\gnuplot=cmr10 at 10pt
\gnuplot
\sbox{\plotpoint}{\rule[-0.200pt]{0.400pt}{0.400pt}}%
\put(181.0,123.0){\rule[-0.200pt]{4.818pt}{0.400pt}}
\put(161,123){\makebox(0,0)[r]{0}}
\put(1419.0,123.0){\rule[-0.200pt]{4.818pt}{0.400pt}}
\put(181.0,246.0){\rule[-0.200pt]{4.818pt}{0.400pt}}
\put(161,246){\makebox(0,0)[r]{500}}
\put(1419.0,246.0){\rule[-0.200pt]{4.818pt}{0.400pt}}
\put(181.0,369.0){\rule[-0.200pt]{4.818pt}{0.400pt}}
\put(161,369){\makebox(0,0)[r]{1000}}
\put(1419.0,369.0){\rule[-0.200pt]{4.818pt}{0.400pt}}
\put(181.0,492.0){\rule[-0.200pt]{4.818pt}{0.400pt}}
\put(161,492){\makebox(0,0)[r]{1500}}
\put(1419.0,492.0){\rule[-0.200pt]{4.818pt}{0.400pt}}
\put(181.0,614.0){\rule[-0.200pt]{4.818pt}{0.400pt}}
\put(161,614){\makebox(0,0)[r]{2000}}
\put(1419.0,614.0){\rule[-0.200pt]{4.818pt}{0.400pt}}
\put(181.0,737.0){\rule[-0.200pt]{4.818pt}{0.400pt}}
\put(161,737){\makebox(0,0)[r]{2500}}
\put(1419.0,737.0){\rule[-0.200pt]{4.818pt}{0.400pt}}
\put(181.0,860.0){\rule[-0.200pt]{4.818pt}{0.400pt}}
\put(161,860){\makebox(0,0)[r]{3000}}
\put(1419.0,860.0){\rule[-0.200pt]{4.818pt}{0.400pt}}
\put(181.0,123.0){\rule[-0.200pt]{0.400pt}{4.818pt}}
\put(181,82){\makebox(0,0){0}}
\put(181.0,840.0){\rule[-0.200pt]{0.400pt}{4.818pt}}
\put(391.0,123.0){\rule[-0.200pt]{0.400pt}{4.818pt}}
\put(391,82){\makebox(0,0){20}}
\put(391.0,840.0){\rule[-0.200pt]{0.400pt}{4.818pt}}
\put(600.0,123.0){\rule[-0.200pt]{0.400pt}{4.818pt}}
\put(600,82){\makebox(0,0){40}}
\put(600.0,840.0){\rule[-0.200pt]{0.400pt}{4.818pt}}
\put(810.0,123.0){\rule[-0.200pt]{0.400pt}{4.818pt}}
\put(810,82){\makebox(0,0){60}}
\put(810.0,840.0){\rule[-0.200pt]{0.400pt}{4.818pt}}
\put(1020.0,123.0){\rule[-0.200pt]{0.400pt}{4.818pt}}
\put(1020,82){\makebox(0,0){80}}
\put(1020.0,840.0){\rule[-0.200pt]{0.400pt}{4.818pt}}
\put(1229.0,123.0){\rule[-0.200pt]{0.400pt}{4.818pt}}
\put(1229,82){\makebox(0,0){100}}
\put(1229.0,840.0){\rule[-0.200pt]{0.400pt}{4.818pt}}
\put(1439.0,123.0){\rule[-0.200pt]{0.400pt}{4.818pt}}
\put(1439,82){\makebox(0,0){120}}
\put(1439.0,840.0){\rule[-0.200pt]{0.400pt}{4.818pt}}
\put(181.0,123.0){\rule[-0.200pt]{303.052pt}{0.400pt}}
\put(1439.0,123.0){\rule[-0.200pt]{0.400pt}{177.543pt}}
\put(181.0,860.0){\rule[-0.200pt]{303.052pt}{0.400pt}}
\put(10,491){\makebox(0,0){$P(T_L)$}}
\put(810,21){\makebox(0,0){$T_L$}}
\put(181.0,123.0){\rule[-0.200pt]{0.400pt}{177.543pt}}
\put(191,123){$\bullet$}
\put(202,123){$\bullet$}
\put(212,123){$\bullet$}
\put(223,123){$\bullet$}
\put(233,123){$\bullet$}
\put(244,123){$\bullet$}
\put(254,123){$\bullet$}
\put(265,123){$\bullet$}
\put(275,123){$\bullet$}
\put(286,123){$\bullet$}
\put(296,124){$\bullet$}
\put(307,123){$\bullet$}
\put(317,124){$\bullet$}
\put(328,124){$\bullet$}
\put(338,125){$\bullet$}
\put(349,126){$\bullet$}
\put(359,126){$\bullet$}
\put(370,128){$\bullet$}
\put(380,126){$\bullet$}
\put(391,128){$\bullet$}
\put(401,129){$\bullet$}
\put(412,134){$\bullet$}
\put(422,135){$\bullet$}
\put(433,138){$\bullet$}
\put(443,136){$\bullet$}
\put(454,140){$\bullet$}
\put(464,141){$\bullet$}
\put(475,141){$\bullet$}
\put(485,148){$\bullet$}
\put(495,149){$\bullet$}
\put(506,157){$\bullet$}
\put(516,161){$\bullet$}
\put(527,173){$\bullet$}
\put(537,171){$\bullet$}
\put(548,179){$\bullet$}
\put(558,190){$\bullet$}
\put(569,191){$\bullet$}
\put(579,204){$\bullet$}
\put(590,213){$\bullet$}
\put(600,216){$\bullet$}
\put(611,231){$\bullet$}
\put(621,241){$\bullet$}
\put(632,255){$\bullet$}
\put(642,262){$\bullet$}
\put(653,277){$\bullet$}
\put(663,293){$\bullet$}
\put(674,297){$\bullet$}
\put(684,324){$\bullet$}
\put(695,328){$\bullet$}
\put(705,355){$\bullet$}
\put(716,374){$\bullet$}
\put(726,381){$\bullet$}
\put(737,400){$\bullet$}
\put(747,439){$\bullet$}
\put(758,440){$\bullet$}
\put(768,459){$\bullet$}
\put(779,512){$\bullet$}
\put(789,516){$\bullet$}
\put(800,541){$\bullet$}
\put(810,553){$\bullet$}
\put(820,596){$\bullet$}
\put(831,608){$\bullet$}
\put(841,625){$\bullet$}
\put(852,663){$\bullet$}
\put(862,682){$\bullet$}
\put(873,709){$\bullet$}
\put(883,739){$\bullet$}
\put(894,756){$\bullet$}
\put(904,768){$\bullet$}
\put(915,792){$\bullet$}
\put(925,810){$\bullet$}
\put(936,794){$\bullet$}
\put(946,808){$\bullet$}
\put(957,841){$\bullet$}
\put(967,823){$\bullet$}
\put(978,832){$\bullet$}
\put(988,811){$\bullet$}
\put(999,814){$\bullet$}
\put(1009,787){$\bullet$}
\put(1020,760){$\bullet$}
\put(1030,743){$\bullet$}
\put(1041,726){$\bullet$}
\put(1051,718){$\bullet$}
\put(1062,664){$\bullet$}
\put(1072,633){$\bullet$}
\put(1083,605){$\bullet$}
\put(1093,560){$\bullet$}
\put(1104,523){$\bullet$}
\put(1114,474){$\bullet$}
\put(1124,443){$\bullet$}
\put(1135,373){$\bullet$}
\put(1145,354){$\bullet$}
\put(1156,317){$\bullet$}
\put(1166,283){$\bullet$}
\put(1177,248){$\bullet$}
\put(1187,222){$\bullet$}
\put(1198,204){$\bullet$}
\put(1208,186){$\bullet$}
\put(1219,175){$\bullet$}
\put(1229,167){$\bullet$}
\put(1240,149){$\bullet$}
\put(1250,139){$\bullet$}
\put(1261,136){$\bullet$}
\put(1271,134){$\bullet$}
\put(1282,128){$\bullet$}
\put(1292,126){$\bullet$}
\put(1303,125){$\bullet$}
\put(1313,123){$\bullet$}
\put(1324,123){$\bullet$}
\put(1334,124){$\bullet$}
\put(1345,123){$\bullet$}
\put(1355,123){$\bullet$}
\put(1366,123){$\bullet$}
\put(1376,123){$\bullet$}
\put(1387,123){$\bullet$}
\put(1397,123){$\bullet$}
\put(1408,123){$\bullet$}
\put(1418,123){$\bullet$}
\put(1429,123){$\bullet$}
\put(1439,123){$\bullet$}
\put(191,123){\usebox{\plotpoint}}
\put(286,122.67){\rule{2.409pt}{0.400pt}}
\multiput(286.00,122.17)(5.000,1.000){2}{\rule{1.204pt}{0.400pt}}
\put(191.0,123.0){\rule[-0.200pt]{22.885pt}{0.400pt}}
\put(328,123.67){\rule{2.409pt}{0.400pt}}
\multiput(328.00,123.17)(5.000,1.000){2}{\rule{1.204pt}{0.400pt}}
\put(296.0,124.0){\rule[-0.200pt]{7.709pt}{0.400pt}}
\put(349,124.67){\rule{2.409pt}{0.400pt}}
\multiput(349.00,124.17)(5.000,1.000){2}{\rule{1.204pt}{0.400pt}}
\put(359,125.67){\rule{2.650pt}{0.400pt}}
\multiput(359.00,125.17)(5.500,1.000){2}{\rule{1.325pt}{0.400pt}}
\put(370,126.67){\rule{2.409pt}{0.400pt}}
\multiput(370.00,126.17)(5.000,1.000){2}{\rule{1.204pt}{0.400pt}}
\put(380,127.67){\rule{2.650pt}{0.400pt}}
\multiput(380.00,127.17)(5.500,1.000){2}{\rule{1.325pt}{0.400pt}}
\put(391,128.67){\rule{2.409pt}{0.400pt}}
\multiput(391.00,128.17)(5.000,1.000){2}{\rule{1.204pt}{0.400pt}}
\put(401,130.17){\rule{2.300pt}{0.400pt}}
\multiput(401.00,129.17)(6.226,2.000){2}{\rule{1.150pt}{0.400pt}}
\put(412,132.17){\rule{2.100pt}{0.400pt}}
\multiput(412.00,131.17)(5.641,2.000){2}{\rule{1.050pt}{0.400pt}}
\put(422,134.17){\rule{2.300pt}{0.400pt}}
\multiput(422.00,133.17)(6.226,2.000){2}{\rule{1.150pt}{0.400pt}}
\put(433,136.17){\rule{2.100pt}{0.400pt}}
\multiput(433.00,135.17)(5.641,2.000){2}{\rule{1.050pt}{0.400pt}}
\put(443,138.17){\rule{2.300pt}{0.400pt}}
\multiput(443.00,137.17)(6.226,2.000){2}{\rule{1.150pt}{0.400pt}}
\multiput(454.00,140.61)(2.025,0.447){3}{\rule{1.433pt}{0.108pt}}
\multiput(454.00,139.17)(7.025,3.000){2}{\rule{0.717pt}{0.400pt}}
\multiput(464.00,143.61)(2.248,0.447){3}{\rule{1.567pt}{0.108pt}}
\multiput(464.00,142.17)(7.748,3.000){2}{\rule{0.783pt}{0.400pt}}
\multiput(475.00,146.60)(1.358,0.468){5}{\rule{1.100pt}{0.113pt}}
\multiput(475.00,145.17)(7.717,4.000){2}{\rule{0.550pt}{0.400pt}}
\multiput(485.00,150.60)(1.358,0.468){5}{\rule{1.100pt}{0.113pt}}
\multiput(485.00,149.17)(7.717,4.000){2}{\rule{0.550pt}{0.400pt}}
\multiput(495.00,154.60)(1.505,0.468){5}{\rule{1.200pt}{0.113pt}}
\multiput(495.00,153.17)(8.509,4.000){2}{\rule{0.600pt}{0.400pt}}
\multiput(506.00,158.59)(1.044,0.477){7}{\rule{0.900pt}{0.115pt}}
\multiput(506.00,157.17)(8.132,5.000){2}{\rule{0.450pt}{0.400pt}}
\multiput(516.00,163.59)(1.155,0.477){7}{\rule{0.980pt}{0.115pt}}
\multiput(516.00,162.17)(8.966,5.000){2}{\rule{0.490pt}{0.400pt}}
\multiput(527.00,168.59)(1.044,0.477){7}{\rule{0.900pt}{0.115pt}}
\multiput(527.00,167.17)(8.132,5.000){2}{\rule{0.450pt}{0.400pt}}
\multiput(537.00,173.59)(0.798,0.485){11}{\rule{0.729pt}{0.117pt}}
\multiput(537.00,172.17)(9.488,7.000){2}{\rule{0.364pt}{0.400pt}}
\multiput(548.00,180.59)(0.852,0.482){9}{\rule{0.767pt}{0.116pt}}
\multiput(548.00,179.17)(8.409,6.000){2}{\rule{0.383pt}{0.400pt}}
\multiput(558.00,186.59)(0.692,0.488){13}{\rule{0.650pt}{0.117pt}}
\multiput(558.00,185.17)(9.651,8.000){2}{\rule{0.325pt}{0.400pt}}
\multiput(569.00,194.59)(0.626,0.488){13}{\rule{0.600pt}{0.117pt}}
\multiput(569.00,193.17)(8.755,8.000){2}{\rule{0.300pt}{0.400pt}}
\multiput(579.00,202.59)(0.692,0.488){13}{\rule{0.650pt}{0.117pt}}
\multiput(579.00,201.17)(9.651,8.000){2}{\rule{0.325pt}{0.400pt}}
\multiput(590.00,210.59)(0.553,0.489){15}{\rule{0.544pt}{0.118pt}}
\multiput(590.00,209.17)(8.870,9.000){2}{\rule{0.272pt}{0.400pt}}
\multiput(600.00,219.58)(0.547,0.491){17}{\rule{0.540pt}{0.118pt}}
\multiput(600.00,218.17)(9.879,10.000){2}{\rule{0.270pt}{0.400pt}}
\multiput(611.58,229.00)(0.491,0.547){17}{\rule{0.118pt}{0.540pt}}
\multiput(610.17,229.00)(10.000,9.879){2}{\rule{0.400pt}{0.270pt}}
\multiput(621.00,240.58)(0.496,0.492){19}{\rule{0.500pt}{0.118pt}}
\multiput(621.00,239.17)(9.962,11.000){2}{\rule{0.250pt}{0.400pt}}
\multiput(632.58,251.00)(0.491,0.600){17}{\rule{0.118pt}{0.580pt}}
\multiput(631.17,251.00)(10.000,10.796){2}{\rule{0.400pt}{0.290pt}}
\multiput(642.58,263.00)(0.492,0.590){19}{\rule{0.118pt}{0.573pt}}
\multiput(641.17,263.00)(11.000,11.811){2}{\rule{0.400pt}{0.286pt}}
\multiput(653.58,276.00)(0.491,0.704){17}{\rule{0.118pt}{0.660pt}}
\multiput(652.17,276.00)(10.000,12.630){2}{\rule{0.400pt}{0.330pt}}
\multiput(663.58,290.00)(0.492,0.684){19}{\rule{0.118pt}{0.645pt}}
\multiput(662.17,290.00)(11.000,13.660){2}{\rule{0.400pt}{0.323pt}}
\multiput(674.58,305.00)(0.491,0.756){17}{\rule{0.118pt}{0.700pt}}
\multiput(673.17,305.00)(10.000,13.547){2}{\rule{0.400pt}{0.350pt}}
\multiput(684.58,320.00)(0.492,0.732){19}{\rule{0.118pt}{0.682pt}}
\multiput(683.17,320.00)(11.000,14.585){2}{\rule{0.400pt}{0.341pt}}
\multiput(695.58,336.00)(0.491,0.912){17}{\rule{0.118pt}{0.820pt}}
\multiput(694.17,336.00)(10.000,16.298){2}{\rule{0.400pt}{0.410pt}}
\multiput(705.58,354.00)(0.492,0.826){19}{\rule{0.118pt}{0.755pt}}
\multiput(704.17,354.00)(11.000,16.434){2}{\rule{0.400pt}{0.377pt}}
\multiput(716.58,372.00)(0.491,0.912){17}{\rule{0.118pt}{0.820pt}}
\multiput(715.17,372.00)(10.000,16.298){2}{\rule{0.400pt}{0.410pt}}
\multiput(726.58,390.00)(0.492,0.920){19}{\rule{0.118pt}{0.827pt}}
\multiput(725.17,390.00)(11.000,18.283){2}{\rule{0.400pt}{0.414pt}}
\multiput(737.58,410.00)(0.491,1.017){17}{\rule{0.118pt}{0.900pt}}
\multiput(736.17,410.00)(10.000,18.132){2}{\rule{0.400pt}{0.450pt}}
\multiput(747.58,430.00)(0.492,1.015){19}{\rule{0.118pt}{0.900pt}}
\multiput(746.17,430.00)(11.000,20.132){2}{\rule{0.400pt}{0.450pt}}
\multiput(758.58,452.00)(0.491,1.121){17}{\rule{0.118pt}{0.980pt}}
\multiput(757.17,452.00)(10.000,19.966){2}{\rule{0.400pt}{0.490pt}}
\multiput(768.58,474.00)(0.492,1.015){19}{\rule{0.118pt}{0.900pt}}
\multiput(767.17,474.00)(11.000,20.132){2}{\rule{0.400pt}{0.450pt}}
\multiput(779.58,496.00)(0.491,1.173){17}{\rule{0.118pt}{1.020pt}}
\multiput(778.17,496.00)(10.000,20.883){2}{\rule{0.400pt}{0.510pt}}
\multiput(789.58,519.00)(0.492,1.109){19}{\rule{0.118pt}{0.973pt}}
\multiput(788.17,519.00)(11.000,21.981){2}{\rule{0.400pt}{0.486pt}}
\multiput(800.58,543.00)(0.491,1.225){17}{\rule{0.118pt}{1.060pt}}
\multiput(799.17,543.00)(10.000,21.800){2}{\rule{0.400pt}{0.530pt}}
\multiput(810.58,567.00)(0.491,1.225){17}{\rule{0.118pt}{1.060pt}}
\multiput(809.17,567.00)(10.000,21.800){2}{\rule{0.400pt}{0.530pt}}
\multiput(820.58,591.00)(0.492,1.109){19}{\rule{0.118pt}{0.973pt}}
\multiput(819.17,591.00)(11.000,21.981){2}{\rule{0.400pt}{0.486pt}}
\multiput(831.58,615.00)(0.491,1.225){17}{\rule{0.118pt}{1.060pt}}
\multiput(830.17,615.00)(10.000,21.800){2}{\rule{0.400pt}{0.530pt}}
\multiput(841.58,639.00)(0.492,1.109){19}{\rule{0.118pt}{0.973pt}}
\multiput(840.17,639.00)(11.000,21.981){2}{\rule{0.400pt}{0.486pt}}
\multiput(852.58,663.00)(0.491,1.173){17}{\rule{0.118pt}{1.020pt}}
\multiput(851.17,663.00)(10.000,20.883){2}{\rule{0.400pt}{0.510pt}}
\multiput(862.58,686.00)(0.492,1.015){19}{\rule{0.118pt}{0.900pt}}
\multiput(861.17,686.00)(11.000,20.132){2}{\rule{0.400pt}{0.450pt}}
\multiput(873.58,708.00)(0.491,1.121){17}{\rule{0.118pt}{0.980pt}}
\multiput(872.17,708.00)(10.000,19.966){2}{\rule{0.400pt}{0.490pt}}
\multiput(883.58,730.00)(0.492,0.873){19}{\rule{0.118pt}{0.791pt}}
\multiput(882.17,730.00)(11.000,17.358){2}{\rule{0.400pt}{0.395pt}}
\multiput(894.58,749.00)(0.491,0.964){17}{\rule{0.118pt}{0.860pt}}
\multiput(893.17,749.00)(10.000,17.215){2}{\rule{0.400pt}{0.430pt}}
\multiput(904.58,768.00)(0.492,0.732){19}{\rule{0.118pt}{0.682pt}}
\multiput(903.17,768.00)(11.000,14.585){2}{\rule{0.400pt}{0.341pt}}
\multiput(915.58,784.00)(0.491,0.704){17}{\rule{0.118pt}{0.660pt}}
\multiput(914.17,784.00)(10.000,12.630){2}{\rule{0.400pt}{0.330pt}}
\multiput(925.58,798.00)(0.492,0.543){19}{\rule{0.118pt}{0.536pt}}
\multiput(924.17,798.00)(11.000,10.887){2}{\rule{0.400pt}{0.268pt}}
\multiput(936.00,810.59)(0.626,0.488){13}{\rule{0.600pt}{0.117pt}}
\multiput(936.00,809.17)(8.755,8.000){2}{\rule{0.300pt}{0.400pt}}
\multiput(946.00,818.59)(0.943,0.482){9}{\rule{0.833pt}{0.116pt}}
\multiput(946.00,817.17)(9.270,6.000){2}{\rule{0.417pt}{0.400pt}}
\put(957,823.67){\rule{2.409pt}{0.400pt}}
\multiput(957.00,823.17)(5.000,1.000){2}{\rule{1.204pt}{0.400pt}}
\put(967,823.17){\rule{2.300pt}{0.400pt}}
\multiput(967.00,824.17)(6.226,-2.000){2}{\rule{1.150pt}{0.400pt}}
\multiput(978.00,821.93)(1.044,-0.477){7}{\rule{0.900pt}{0.115pt}}
\multiput(978.00,822.17)(8.132,-5.000){2}{\rule{0.450pt}{0.400pt}}
\multiput(988.00,816.92)(0.547,-0.491){17}{\rule{0.540pt}{0.118pt}}
\multiput(988.00,817.17)(9.879,-10.000){2}{\rule{0.270pt}{0.400pt}}
\multiput(999.58,805.26)(0.491,-0.704){17}{\rule{0.118pt}{0.660pt}}
\multiput(998.17,806.63)(10.000,-12.630){2}{\rule{0.400pt}{0.330pt}}
\multiput(1009.58,790.72)(0.492,-0.873){19}{\rule{0.118pt}{0.791pt}}
\multiput(1008.17,792.36)(11.000,-17.358){2}{\rule{0.400pt}{0.395pt}}
\multiput(1020.58,770.93)(0.491,-1.121){17}{\rule{0.118pt}{0.980pt}}
\multiput(1019.17,772.97)(10.000,-19.966){2}{\rule{0.400pt}{0.490pt}}
\multiput(1030.58,748.66)(0.492,-1.203){19}{\rule{0.118pt}{1.045pt}}
\multiput(1029.17,750.83)(11.000,-23.830){2}{\rule{0.400pt}{0.523pt}}
\multiput(1041.58,721.60)(0.491,-1.538){17}{\rule{0.118pt}{1.300pt}}
\multiput(1040.17,724.30)(10.000,-27.302){2}{\rule{0.400pt}{0.650pt}}
\multiput(1051.58,691.60)(0.492,-1.534){19}{\rule{0.118pt}{1.300pt}}
\multiput(1050.17,694.30)(11.000,-30.302){2}{\rule{0.400pt}{0.650pt}}
\multiput(1062.58,657.61)(0.491,-1.850){17}{\rule{0.118pt}{1.540pt}}
\multiput(1061.17,660.80)(10.000,-32.804){2}{\rule{0.400pt}{0.770pt}}
\multiput(1072.58,621.85)(0.492,-1.769){19}{\rule{0.118pt}{1.482pt}}
\multiput(1071.17,624.92)(11.000,-34.924){2}{\rule{0.400pt}{0.741pt}}
\multiput(1083.58,583.11)(0.491,-2.007){17}{\rule{0.118pt}{1.660pt}}
\multiput(1082.17,586.55)(10.000,-35.555){2}{\rule{0.400pt}{0.830pt}}
\multiput(1093.58,544.40)(0.492,-1.911){19}{\rule{0.118pt}{1.591pt}}
\multiput(1092.17,547.70)(11.000,-37.698){2}{\rule{0.400pt}{0.795pt}}
\multiput(1104.58,502.78)(0.491,-2.111){17}{\rule{0.118pt}{1.740pt}}
\multiput(1103.17,506.39)(10.000,-37.389){2}{\rule{0.400pt}{0.870pt}}
\multiput(1114.58,461.94)(0.491,-2.059){17}{\rule{0.118pt}{1.700pt}}
\multiput(1113.17,465.47)(10.000,-36.472){2}{\rule{0.400pt}{0.850pt}}
\multiput(1124.58,422.70)(0.492,-1.817){19}{\rule{0.118pt}{1.518pt}}
\multiput(1123.17,425.85)(11.000,-35.849){2}{\rule{0.400pt}{0.759pt}}
\multiput(1135.58,383.44)(0.491,-1.903){17}{\rule{0.118pt}{1.580pt}}
\multiput(1134.17,386.72)(10.000,-33.721){2}{\rule{0.400pt}{0.790pt}}
\multiput(1145.58,347.30)(0.492,-1.628){19}{\rule{0.118pt}{1.373pt}}
\multiput(1144.17,350.15)(11.000,-32.151){2}{\rule{0.400pt}{0.686pt}}
\multiput(1156.58,312.11)(0.491,-1.694){17}{\rule{0.118pt}{1.420pt}}
\multiput(1155.17,315.05)(10.000,-30.053){2}{\rule{0.400pt}{0.710pt}}
\multiput(1166.58,280.21)(0.492,-1.345){19}{\rule{0.118pt}{1.155pt}}
\multiput(1165.17,282.60)(11.000,-26.604){2}{\rule{0.400pt}{0.577pt}}
\multiput(1177.58,251.43)(0.491,-1.277){17}{\rule{0.118pt}{1.100pt}}
\multiput(1176.17,253.72)(10.000,-22.717){2}{\rule{0.400pt}{0.550pt}}
\multiput(1187.58,227.11)(0.492,-1.062){19}{\rule{0.118pt}{0.936pt}}
\multiput(1186.17,229.06)(11.000,-21.057){2}{\rule{0.400pt}{0.468pt}}
\multiput(1198.58,204.43)(0.491,-0.964){17}{\rule{0.118pt}{0.860pt}}
\multiput(1197.17,206.22)(10.000,-17.215){2}{\rule{0.400pt}{0.430pt}}
\multiput(1208.58,186.17)(0.492,-0.732){19}{\rule{0.118pt}{0.682pt}}
\multiput(1207.17,187.58)(11.000,-14.585){2}{\rule{0.400pt}{0.341pt}}
\multiput(1219.58,170.59)(0.491,-0.600){17}{\rule{0.118pt}{0.580pt}}
\multiput(1218.17,171.80)(10.000,-10.796){2}{\rule{0.400pt}{0.290pt}}
\multiput(1229.00,159.92)(0.496,-0.492){19}{\rule{0.500pt}{0.118pt}}
\multiput(1229.00,160.17)(9.962,-11.000){2}{\rule{0.250pt}{0.400pt}}
\multiput(1240.00,148.93)(0.626,-0.488){13}{\rule{0.600pt}{0.117pt}}
\multiput(1240.00,149.17)(8.755,-8.000){2}{\rule{0.300pt}{0.400pt}}
\multiput(1250.00,140.93)(0.943,-0.482){9}{\rule{0.833pt}{0.116pt}}
\multiput(1250.00,141.17)(9.270,-6.000){2}{\rule{0.417pt}{0.400pt}}
\multiput(1261.00,134.94)(1.358,-0.468){5}{\rule{1.100pt}{0.113pt}}
\multiput(1261.00,135.17)(7.717,-4.000){2}{\rule{0.550pt}{0.400pt}}
\multiput(1271.00,130.95)(2.248,-0.447){3}{\rule{1.567pt}{0.108pt}}
\multiput(1271.00,131.17)(7.748,-3.000){2}{\rule{0.783pt}{0.400pt}}
\put(1282,127.17){\rule{2.100pt}{0.400pt}}
\multiput(1282.00,128.17)(5.641,-2.000){2}{\rule{1.050pt}{0.400pt}}
\put(1292,125.17){\rule{2.300pt}{0.400pt}}
\multiput(1292.00,126.17)(6.226,-2.000){2}{\rule{1.150pt}{0.400pt}}
\put(1303,123.67){\rule{2.409pt}{0.400pt}}
\multiput(1303.00,124.17)(5.000,-1.000){2}{\rule{1.204pt}{0.400pt}}
\put(338.0,125.0){\rule[-0.200pt]{2.650pt}{0.400pt}}
\put(1324,122.67){\rule{2.409pt}{0.400pt}}
\multiput(1324.00,123.17)(5.000,-1.000){2}{\rule{1.204pt}{0.400pt}}
\put(1313.0,124.0){\rule[-0.200pt]{2.650pt}{0.400pt}}
\put(1334.0,123.0){\rule[-0.200pt]{25.294pt}{0.400pt}}
\put(191,128){\usebox{\plotpoint}}
\multiput(223.00,128.61)(2.025,0.447){3}{\rule{1.433pt}{0.108pt}}
\multiput(223.00,127.17)(7.025,3.000){2}{\rule{0.717pt}{0.400pt}}
\put(191.0,128.0){\rule[-0.200pt]{7.709pt}{0.400pt}}
\multiput(328.58,131.00)(0.491,1.121){17}{\rule{0.118pt}{0.980pt}}
\multiput(327.17,131.00)(10.000,19.966){2}{\rule{0.400pt}{0.490pt}}
\put(233.0,131.0){\rule[-0.200pt]{22.885pt}{0.400pt}}
\multiput(433.58,153.00)(0.491,5.134){17}{\rule{0.118pt}{4.060pt}}
\multiput(432.17,153.00)(10.000,90.573){2}{\rule{0.400pt}{2.030pt}}
\put(338.0,153.0){\rule[-0.200pt]{22.885pt}{0.400pt}}
\multiput(642.58,252.00)(0.492,12.099){19}{\rule{0.118pt}{9.445pt}}
\multiput(641.17,252.00)(11.000,237.395){2}{\rule{0.400pt}{4.723pt}}
\put(443.0,252.0){\rule[-0.200pt]{47.939pt}{0.400pt}}
\multiput(852.58,509.00)(0.491,3.101){17}{\rule{0.118pt}{2.500pt}}
\multiput(851.17,509.00)(10.000,54.811){2}{\rule{0.400pt}{1.250pt}}
\put(653.0,509.0){\rule[-0.200pt]{47.939pt}{0.400pt}}
\multiput(957.58,569.00)(0.491,12.536){17}{\rule{0.118pt}{9.740pt}}
\multiput(956.17,569.00)(10.000,220.784){2}{\rule{0.400pt}{4.870pt}}
\put(862.0,569.0){\rule[-0.200pt]{22.885pt}{0.400pt}}
\multiput(1062.58,800.95)(0.491,-2.684){17}{\rule{0.118pt}{2.180pt}}
\multiput(1061.17,805.48)(10.000,-47.475){2}{\rule{0.400pt}{1.090pt}}
\put(967.0,810.0){\rule[-0.200pt]{22.885pt}{0.400pt}}
\put(1072.0,758.0){\rule[-0.200pt]{37.821pt}{0.400pt}}
\end{picture}

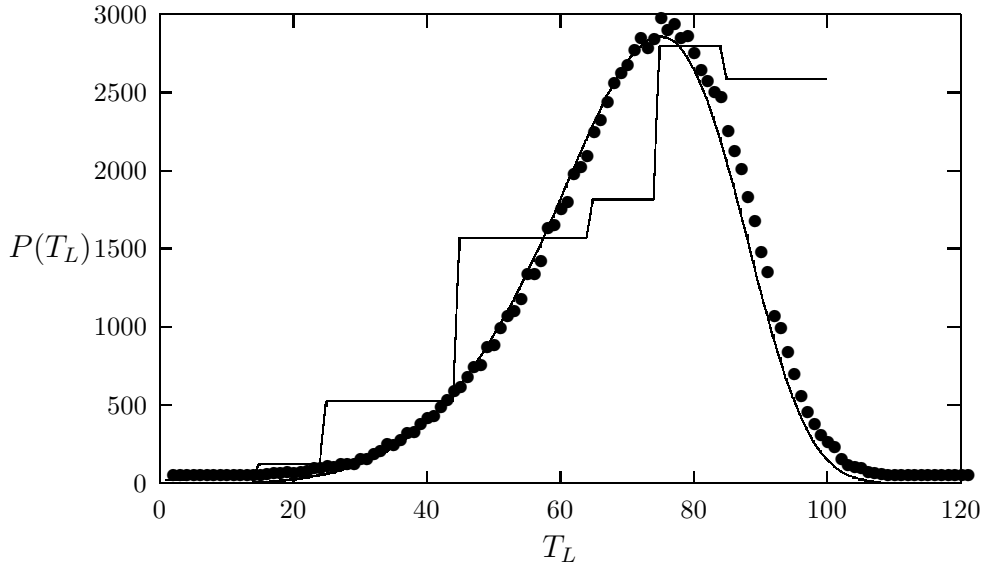
\noindent {\small Fig.1. The unnormalised distribution of lifetimes 
(used in the simulation) is represented by $\bullet$ symbol
(generated by Monte Carlo method for $N = 100000$). 
The solid smooth curve indicates the functional
form of the distribution function $P(x)=Ax^a{\rm exp}(-{x^b}/c)$, 
taking $a=4.0$, $b=7.5$ and $c=2.16\times10^{14}$. The histogram 
drawn by using the data given in table-I.}

\vskip 0.5 cm

We studied the net gain $G_n/N$ (per head) as a function of $T_f$. 
We found the upper age (upto which a person is liable to pay the premium) 
$T_f$ for which the IC can get
maximum gain. We found it is at 57 years. This result is plausible
% consistent ==> plausible (twice)
 since it will be below the most probable lifetime $T_m$. Another important
thing is the upper age limit for which the gain is negative i.e. loss. This we find 
(in fig-2) is around 77 years. It is also plausible
since it is above the most probable lifetime ($T_m$=75 yrs). If any IC sets $T_f$ 
much above $T_m$ it definitely has to pay lots of money for deaths
and cannot make the profit.

From these results, we observed that the upper age for maximum gain (per head) and the age for 
loss are independent of number of policy holders. This was shown in fig-2. The value of
maximum gain per head is also independent of the number of policy holders. We have studied this 
for $N$ = 1000, 10000 and 100000. For $N=1000$,
the gain is slightly lower that that obtained for $N$ = 10000 and 100000. Figure-2 shows that 
the data for $N$ = 10000 and that for $N$ = 100000 
are marginally distinguishable. That means the data for $N = 100000$ do not suffer from finite size effect.

\vskip 2 cm

%%%% FIGURE-2
% GNUPLOT: LaTeX picture
\setlength{\unitlength}{0.240900pt}
\ifx\plotpoint\undefined\newsavebox{\plotpoint}\fi
\sbox{\plotpoint}{\rule[-0.200pt]{0.400pt}{0.400pt}}%
\begin{picture}(1500,900)(0,0)
\font\gnuplot=cmr10 at 10pt
\gnuplot
\sbox{\plotpoint}{\rule[-0.200pt]{0.400pt}{0.400pt}}%
\put(181.0,123.0){\rule[-0.200pt]{4.818pt}{0.400pt}}
\put(161,123){\makebox(0,0)[r]{-0.4}}
\put(1419.0,123.0){\rule[-0.200pt]{4.818pt}{0.400pt}}
\put(181.0,246.0){\rule[-0.200pt]{4.818pt}{0.400pt}}
\put(161,246){\makebox(0,0)[r]{-0.2}}
\put(1419.0,246.0){\rule[-0.200pt]{4.818pt}{0.400pt}}
\put(181.0,369.0){\rule[-0.200pt]{4.818pt}{0.400pt}}
\put(161,369){\makebox(0,0)[r]{0}}
\put(1419.0,369.0){\rule[-0.200pt]{4.818pt}{0.400pt}}
\put(181.0,491.0){\rule[-0.200pt]{4.818pt}{0.400pt}}
\put(161,491){\makebox(0,0)[r]{0.2}}
\put(1419.0,491.0){\rule[-0.200pt]{4.818pt}{0.400pt}}
\put(181.0,614.0){\rule[-0.200pt]{4.818pt}{0.400pt}}
\put(161,614){\makebox(0,0)[r]{0.4}}
\put(1419.0,614.0){\rule[-0.200pt]{4.818pt}{0.400pt}}
\put(181.0,737.0){\rule[-0.200pt]{4.818pt}{0.400pt}}
\put(161,737){\makebox(0,0)[r]{0.6}}
\put(1419.0,737.0){\rule[-0.200pt]{4.818pt}{0.400pt}}
\put(181.0,860.0){\rule[-0.200pt]{4.818pt}{0.400pt}}
\put(161,860){\makebox(0,0)[r]{0.8}}
\put(1419.0,860.0){\rule[-0.200pt]{4.818pt}{0.400pt}}
\put(181.0,123.0){\rule[-0.200pt]{0.400pt}{4.818pt}}
\put(181,82){\makebox(0,0){40}}
\put(181.0,840.0){\rule[-0.200pt]{0.400pt}{4.818pt}}
\put(338.0,123.0){\rule[-0.200pt]{0.400pt}{4.818pt}}
\put(338,82){\makebox(0,0){45}}
\put(338.0,840.0){\rule[-0.200pt]{0.400pt}{4.818pt}}
\put(495.0,123.0){\rule[-0.200pt]{0.400pt}{4.818pt}}
\put(495,82){\makebox(0,0){50}}
\put(495.0,840.0){\rule[-0.200pt]{0.400pt}{4.818pt}}
\put(653.0,123.0){\rule[-0.200pt]{0.400pt}{4.818pt}}
\put(653,82){\makebox(0,0){55}}
\put(653.0,840.0){\rule[-0.200pt]{0.400pt}{4.818pt}}
\put(810.0,123.0){\rule[-0.200pt]{0.400pt}{4.818pt}}
\put(810,82){\makebox(0,0){60}}
\put(810.0,840.0){\rule[-0.200pt]{0.400pt}{4.818pt}}
\put(967.0,123.0){\rule[-0.200pt]{0.400pt}{4.818pt}}
\put(967,82){\makebox(0,0){65}}
\put(967.0,840.0){\rule[-0.200pt]{0.400pt}{4.818pt}}
\put(1124.0,123.0){\rule[-0.200pt]{0.400pt}{4.818pt}}
\put(1124,82){\makebox(0,0){70}}
\put(1124.0,840.0){\rule[-0.200pt]{0.400pt}{4.818pt}}
\put(1282.0,123.0){\rule[-0.200pt]{0.400pt}{4.818pt}}
\put(1282,82){\makebox(0,0){75}}
\put(1282.0,840.0){\rule[-0.200pt]{0.400pt}{4.818pt}}
\put(1439.0,123.0){\rule[-0.200pt]{0.400pt}{4.818pt}}
\put(1439,82){\makebox(0,0){80}}
\put(1439.0,840.0){\rule[-0.200pt]{0.400pt}{4.818pt}}
\put(181.0,123.0){\rule[-0.200pt]{303.052pt}{0.400pt}}
\put(1439.0,123.0){\rule[-0.200pt]{0.400pt}{177.543pt}}
\put(181.0,860.0){\rule[-0.200pt]{303.052pt}{0.400pt}}
\put(40,491){\makebox(0,0){${{G_n} \over N}$}}
\put(810,21){\makebox(0,0){$T_f$}}
\put(181.0,123.0){\rule[-0.200pt]{0.400pt}{177.543pt}}
\put(181,669){\raisebox{-.8pt}{\makebox(0,0){$\Diamond$}}}
\put(212,680){\raisebox{-.8pt}{\makebox(0,0){$\Diamond$}}}
\put(244,692){\raisebox{-.8pt}{\makebox(0,0){$\Diamond$}}}
\put(275,704){\raisebox{-.8pt}{\makebox(0,0){$\Diamond$}}}
\put(307,716){\raisebox{-.8pt}{\makebox(0,0){$\Diamond$}}}
\put(338,727){\raisebox{-.8pt}{\makebox(0,0){$\Diamond$}}}
\put(370,738){\raisebox{-.8pt}{\makebox(0,0){$\Diamond$}}}
\put(401,748){\raisebox{-.8pt}{\makebox(0,0){$\Diamond$}}}
\put(433,758){\raisebox{-.8pt}{\makebox(0,0){$\Diamond$}}}
\put(464,767){\raisebox{-.8pt}{\makebox(0,0){$\Diamond$}}}
\put(495,776){\raisebox{-.8pt}{\makebox(0,0){$\Diamond$}}}
\put(527,784){\raisebox{-.8pt}{\makebox(0,0){$\Diamond$}}}
\put(558,790){\raisebox{-.8pt}{\makebox(0,0){$\Diamond$}}}
\put(590,796){\raisebox{-.8pt}{\makebox(0,0){$\Diamond$}}}
\put(621,800){\raisebox{-.8pt}{\makebox(0,0){$\Diamond$}}}
\put(653,803){\raisebox{-.8pt}{\makebox(0,0){$\Diamond$}}}
\put(684,805){\raisebox{-.8pt}{\makebox(0,0){$\Diamond$}}}
\put(716,805){\raisebox{-.8pt}{\makebox(0,0){$\Diamond$}}}
\put(747,804){\raisebox{-.8pt}{\makebox(0,0){$\Diamond$}}}
\put(779,801){\raisebox{-.8pt}{\makebox(0,0){$\Diamond$}}}
\put(810,796){\raisebox{-.8pt}{\makebox(0,0){$\Diamond$}}}
\put(841,789){\raisebox{-.8pt}{\makebox(0,0){$\Diamond$}}}
\put(873,780){\raisebox{-.8pt}{\makebox(0,0){$\Diamond$}}}
\put(904,769){\raisebox{-.8pt}{\makebox(0,0){$\Diamond$}}}
\put(936,756){\raisebox{-.8pt}{\makebox(0,0){$\Diamond$}}}
\put(967,740){\raisebox{-.8pt}{\makebox(0,0){$\Diamond$}}}
\put(999,722){\raisebox{-.8pt}{\makebox(0,0){$\Diamond$}}}
\put(1030,702){\raisebox{-.8pt}{\makebox(0,0){$\Diamond$}}}
\put(1062,679){\raisebox{-.8pt}{\makebox(0,0){$\Diamond$}}}
\put(1093,653){\raisebox{-.8pt}{\makebox(0,0){$\Diamond$}}}
\put(1124,624){\raisebox{-.8pt}{\makebox(0,0){$\Diamond$}}}
\put(1156,592){\raisebox{-.8pt}{\makebox(0,0){$\Diamond$}}}
\put(1187,557){\raisebox{-.8pt}{\makebox(0,0){$\Diamond$}}}
\put(1219,520){\raisebox{-.8pt}{\makebox(0,0){$\Diamond$}}}
\put(1250,480){\raisebox{-.8pt}{\makebox(0,0){$\Diamond$}}}
\put(1282,436){\raisebox{-.8pt}{\makebox(0,0){$\Diamond$}}}
\put(1313,390){\raisebox{-.8pt}{\makebox(0,0){$\Diamond$}}}
\put(1345,341){\raisebox{-.8pt}{\makebox(0,0){$\Diamond$}}}
\put(1376,289){\raisebox{-.8pt}{\makebox(0,0){$\Diamond$}}}
\put(1408,234){\raisebox{-.8pt}{\makebox(0,0){$\Diamond$}}}
\put(1439,175){\raisebox{-.8pt}{\makebox(0,0){$\Diamond$}}}
\put(181,683){\makebox(0,0){$+$}}
\put(212,696){\makebox(0,0){$+$}}
\put(244,708){\makebox(0,0){$+$}}
\put(275,720){\makebox(0,0){$+$}}
\put(307,732){\makebox(0,0){$+$}}
\put(338,744){\makebox(0,0){$+$}}
\put(370,756){\makebox(0,0){$+$}}
\put(401,766){\makebox(0,0){$+$}}
\put(433,777){\makebox(0,0){$+$}}
\put(464,786){\makebox(0,0){$+$}}
\put(495,795){\makebox(0,0){$+$}}
\put(527,803){\makebox(0,0){$+$}}
\put(558,810){\makebox(0,0){$+$}}
\put(590,816){\makebox(0,0){$+$}}
\put(621,821){\makebox(0,0){$+$}}
\put(653,824){\makebox(0,0){$+$}}
\put(684,827){\makebox(0,0){$+$}}
\put(716,828){\makebox(0,0){$+$}}
\put(747,827){\makebox(0,0){$+$}}
\put(779,825){\makebox(0,0){$+$}}
\put(810,821){\makebox(0,0){$+$}}
\put(841,815){\makebox(0,0){$+$}}
\put(873,807){\makebox(0,0){$+$}}
\put(904,797){\makebox(0,0){$+$}}
\put(936,785){\makebox(0,0){$+$}}
\put(967,770){\makebox(0,0){$+$}}
\put(999,753){\makebox(0,0){$+$}}
\put(1030,733){\makebox(0,0){$+$}}
\put(1062,710){\makebox(0,0){$+$}}
\put(1093,684){\makebox(0,0){$+$}}
\put(1124,655){\makebox(0,0){$+$}}
\put(1156,623){\makebox(0,0){$+$}}
\put(1187,588){\makebox(0,0){$+$}}
\put(1219,550){\makebox(0,0){$+$}}
\put(1250,508){\makebox(0,0){$+$}}
\put(1282,464){\makebox(0,0){$+$}}
\put(1313,415){\makebox(0,0){$+$}}
\put(1345,364){\makebox(0,0){$+$}}
\put(1376,308){\makebox(0,0){$+$}}
\put(1408,249){\makebox(0,0){$+$}}
\put(1439,186){\makebox(0,0){$+$}}
\sbox{\plotpoint}{\rule[-0.400pt]{0.800pt}{0.800pt}}%
\put(181,685){\raisebox{-.8pt}{\makebox(0,0){$\Box$}}}
\put(212,697){\raisebox{-.8pt}{\makebox(0,0){$\Box$}}}
\put(244,710){\raisebox{-.8pt}{\makebox(0,0){$\Box$}}}
\put(275,722){\raisebox{-.8pt}{\makebox(0,0){$\Box$}}}
\put(307,734){\raisebox{-.8pt}{\makebox(0,0){$\Box$}}}
\put(338,746){\raisebox{-.8pt}{\makebox(0,0){$\Box$}}}
\put(370,757){\raisebox{-.8pt}{\makebox(0,0){$\Box$}}}
\put(401,767){\raisebox{-.8pt}{\makebox(0,0){$\Box$}}}
\put(433,777){\raisebox{-.8pt}{\makebox(0,0){$\Box$}}}
\put(464,786){\raisebox{-.8pt}{\makebox(0,0){$\Box$}}}
\put(495,795){\raisebox{-.8pt}{\makebox(0,0){$\Box$}}}
\put(527,802){\raisebox{-.8pt}{\makebox(0,0){$\Box$}}}
\put(558,809){\raisebox{-.8pt}{\makebox(0,0){$\Box$}}}
\put(590,815){\raisebox{-.8pt}{\makebox(0,0){$\Box$}}}
\put(621,819){\raisebox{-.8pt}{\makebox(0,0){$\Box$}}}
\put(653,823){\raisebox{-.8pt}{\makebox(0,0){$\Box$}}}
\put(684,825){\raisebox{-.8pt}{\makebox(0,0){$\Box$}}}
\put(716,825){\raisebox{-.8pt}{\makebox(0,0){$\Box$}}}
\put(747,824){\raisebox{-.8pt}{\makebox(0,0){$\Box$}}}
\put(779,822){\raisebox{-.8pt}{\makebox(0,0){$\Box$}}}
\put(810,817){\raisebox{-.8pt}{\makebox(0,0){$\Box$}}}
\put(841,811){\raisebox{-.8pt}{\makebox(0,0){$\Box$}}}
\put(873,803){\raisebox{-.8pt}{\makebox(0,0){$\Box$}}}
\put(904,792){\raisebox{-.8pt}{\makebox(0,0){$\Box$}}}
\put(936,780){\raisebox{-.8pt}{\makebox(0,0){$\Box$}}}
\put(967,765){\raisebox{-.8pt}{\makebox(0,0){$\Box$}}}
\put(999,747){\raisebox{-.8pt}{\makebox(0,0){$\Box$}}}
\put(1030,727){\raisebox{-.8pt}{\makebox(0,0){$\Box$}}}
\put(1062,705){\raisebox{-.8pt}{\makebox(0,0){$\Box$}}}
\put(1093,679){\raisebox{-.8pt}{\makebox(0,0){$\Box$}}}
\put(1124,651){\raisebox{-.8pt}{\makebox(0,0){$\Box$}}}
\put(1156,619){\raisebox{-.8pt}{\makebox(0,0){$\Box$}}}
\put(1187,584){\raisebox{-.8pt}{\makebox(0,0){$\Box$}}}
\put(1219,546){\raisebox{-.8pt}{\makebox(0,0){$\Box$}}}
\put(1250,505){\raisebox{-.8pt}{\makebox(0,0){$\Box$}}}
\put(1282,460){\raisebox{-.8pt}{\makebox(0,0){$\Box$}}}
\put(1313,412){\raisebox{-.8pt}{\makebox(0,0){$\Box$}}}
\put(1345,360){\raisebox{-.8pt}{\makebox(0,0){$\Box$}}}
\put(1376,305){\raisebox{-.8pt}{\makebox(0,0){$\Box$}}}
\put(1408,246){\raisebox{-.8pt}{\makebox(0,0){$\Box$}}}
\put(1439,184){\raisebox{-.8pt}{\makebox(0,0){$\Box$}}}
\end{picture}

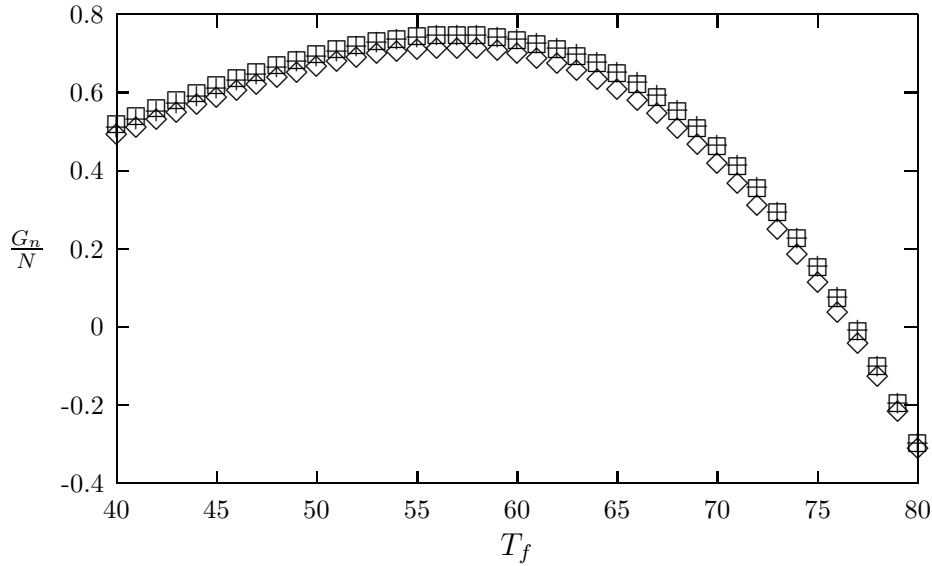
\noindent {\small Fig.2. The variation of $G_n/N$ with respect to $T_f$. 
The curve shows $G_n/N$ will be maximum for $T_f=57$ and it
will be negative for $T_f=77$. Different symbols represent different number 
of persons. $N = 1000 (\Diamond)$, $N = 10000 (+)$
and $N = 100000 (\Box)$.} 

\vskip 1cm

\noindent {\bf 4. Summary}

We proposed a model insurance (life) policy and studied it by computer simulation. 
For large number of policy holders
($N = 100000$), we have considered a nonmonotonic statistical distribution of 
lifetimes. The insurance company came
to the market and give insurance to all people having same age. We studied the 
net gain (per head) of IC as a function 
of the upper age of last premium. Interestingly, we found that the gain becomes 
maximum for a particular upper age
of last premium. For this results the IC can get some idea about the upper age 
of last premium. Another important
thing is the upper age limit of last premium for {\it marginal} profit. We observed 
that if $T_f$ is quite high, so that
a large number of policy holders die before their age of last premium the company can 
not get profit. This information
is also important to the IC.

We have some plan to study the net gain as function of $T_i$. There are several important 
studies have to be done.
The dependence of net gain on the rates of interest $r_1$ and $r_2$. 
The model is a simplified model. One has to consider the other
factors like, laps of policy, the various values of sum assured for different 
types of death (accidental, natural etc.).
It will also be important to know how this profit depends on the half width of 
the lifetime distribution. 
We are trying to modify the model considering these factors to make it as realistic 
as possible. Our work is going 
on and the details will be reported elsewhere.
 
\vskip 1cm

\noindent {\bf Acknowledgments:}
The library facility provided by Saha Institute of Nuclear Physics, 
Calcutta, is gratefully acknowledged.
We would like to thank the referee for bringing ref.\cite{mort} into our 
notice and for important
suggestions.

\vskip 1cm

\noindent {\large {\bf References}}

\end{document}